\documentclass[conference,letterpaper]{IEEEtran}

\usepackage[utf8]{inputenc} 
\usepackage[T1]{fontenc}
\usepackage{url}
\usepackage{ifthen}
\usepackage{cite}
\usepackage[cmex10]{amsmath} 

\usepackage{graphicx}         
\usepackage{booktabs}         
\usepackage{amsmath,amsfonts,bm}

\usepackage{mathrsfs}

\usepackage{algorithm}
\usepackage{algorithmic}
\usepackage{array}
\usepackage{textcomp}
\usepackage{stfloats}
\usepackage{url}
\usepackage{verbatim}
\usepackage{amsmath,amssymb,amsfonts}
\usepackage{cite}
\usepackage{pgf}
\usepackage{tikz}
\usetikzlibrary{shapes,arrows,decorations.markings}
\usetikzlibrary{positioning,arrows.meta}
\usetikzlibrary{positioning}
\usepackage{xcolor}
\usepackage{smartdiagram}
\usepackage{amsmath}
\usepackage{caption}
\usepackage{subcaption}
\begin{document}
\title{DeepPolar+: Breaking the BER-BLER Trade-off with Self-Attention and SMART (SNR-MAtched Redundancy Technique) decoding} 


\author{
	\IEEEauthorblockN{ Shubham Srivastava, and Adrish Banerjee} 
	\IEEEauthorblockA{ Department of Electrical Engineering, Indian Institute of Technology Kanpur, India \\
		Email: \{shubhsr, adrish\}@iitk.ac.in           
	}
}

\maketitle


\begin{abstract}
	DeepPolar codes have recently emerged as a promising approach for channel coding, demonstrating superior bit error rate (BER) performance compared to conventional polar codes. Despite their excellent BER characteristics, these codes exhibit suboptimal block error rate (BLER) performance, creating a fundamental BER-BLER trade-off that severely limits their practical deployment in communication systems. This paper introduces DeepPolar+, an enhanced neural polar coding framework that systematically eliminates this BER-BLER trade-off by simultaneously improving BLER performance while maintaining the superior BER characteristics of DeepPolar codes. Our approach achieves this breakthrough through three key innovations: (1) an attention-enhanced decoder architecture that leverages multi-head self-attention mechanisms to capture complex dependencies between bit positions, (2) a structured loss function that jointly optimizes for both bit-level accuracy and block-level reliability, and (3) an adaptive SNR-Matched Redundancy Technique (SMART) for decoding DeepPolar+ code (DP+SMART decoder) that combines specialized models with CRC verification for robust performance across diverse channel conditions. For a (256,37) code configuration, DeepPolar+ demonstrates notable improvements in both BER and BLER performance compared to conventional successive cancellation decoding and DeepPolar, while achieving remarkably faster convergence through improved architecture and optimization strategies. The DeepPolar+SMART variant further amplifies these dual improvements, delivering significant gains in both error rate metrics over existing approaches. DeepPolar+ effectively bridges the gap between theoretical potential and practical implementation of neural polar codes, offering a viable path forward for next-generation error correction systems.
\end{abstract}

\section{Introduction}\label{Sec: intro}

\subsection{Evolution of Channel Codes}
Channel coding lies at the heart of reliable digital communication. From Reed-Muller codes \cite{1057465}, \cite{6499441} to Turbo codes \cite{397441} and LDPC codes \cite{gallager1963low}, each advancement has shaped modern communication systems. Polar codes, introduced by Arıkan \cite{4595172}, represent a breakthrough as the first explicitly constructed capacity-achieving codes with efficient encoding and decoding algorithms. Their swift integration into 5G standards \cite{3GPP2018, bioglio2020design, 8477009} within a decade underscores their practical significance.

\subsection{Recent Advances in Polar Codes}
While Polar codes achieve  Shannon capacity asymptotically through successive cancellation (SC) decoding, their finite-length performance remains suboptimal. Some of the recent enhancements \cite{niu2012crc, wang2016parity, trifonov2013polar, trifonov2015polar, zhang2018parity, arikan2019sequential} include CRC-aided SCL decoding \cite{tal2015list, niu2012crc} and polarization-adjusted convolutional (PAC) codes \cite{arikan2019sequential}. Another promising direction explores larger kernel dimensions. Korada et al. \cite{korada2010polar} proved that polarization occurs for kernels of any size, provided they are non-unitary and not upper triangular under column permutations. However, practical implementation of large-kernel polar codes faces computational challenges in decoding.


\subsection{Deep Learning in Channel Coding}
The intersection of deep learning and channel coding has yielded significant advances in error correction systems \cite{Hebbar_2022, 9174106}. While initial efforts focused on enhancing existing decoders \cite{7852251, dorner2017deep, nachmani2019hyper}, designing completely new codes through deep learning proved challenging. A breakthrough came through neural architectures that preserve coding-theoretic structures while introducing learned components, as demonstrated in Turbo codes \cite{jiang2019turbo}, KO codes \cite{pmlr-v139-makkuva21a}, and ProductAE \cite{vahid2021productae}. Recent work like Dense KO (DKO) codes \cite{10619329} has further improved efficiency through architectural innovations.
\subsection{Deep Learning for Polar Codes}
Deep learning applications to Polar codes have evolved from decoder enhancement \cite{cammerer2017combining, hebbar2023crisp} and frozen bit selection \cite{liao2021construction, ankireddy2024nested} to complete architectural innovations. DeepPolar codes \cite{hebbar2024deeppolar} represent a significant advance, introducing learnable non-linear generalizations of the polarization kernel while maintaining the essential structural benefits of conventional polar codes. However, a fundamental limitation emerges in the trade-off between BER and BLER performance, hindering practical adoption.

\subsection{Our Contributions}
This paper introduces DeepPolar+, which resolves this limitation through three key innovations:

\begin{itemize}
	\item An attention-enhanced decoder architecture leveraging multi-head self-attention mechanisms to capture complex dependencies between bit positions
	\item A structured loss function jointly optimizing bit-level accuracy and block-level reliability
	\item An adaptive SNR-Matched Redundancy Technique for decoding DeepPolar+ code.
\end{itemize}


\subsection{Code Availability and Organization}
Our implementation is publicly available at: \url{https://github.com/shubhamsrivast4u/DeepPolarPlus}

The rest of the paper is organized as follows: Section \ref{Sec: background} provides background on DeepPolar codes. Section \ref{Sec: deepolar+} details the DeepPolar+ architecture and training methodology. Section \ref{Sec:results} presents experimental results and analysis. Section \ref{Sec: conclusion} concludes the paper.

\begin{figure*}[t]
	\centering
	\begin{subfigure}[t]{0.49\textwidth} 
			\centering
		\includegraphics[width=1\textwidth]{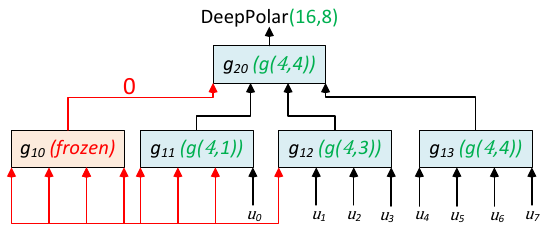}
		\caption{}
		\label{fig:encodertree}
	\end{subfigure}
	\hfill
	\centering
	\begin{subfigure}[t]{0.49\textwidth} 
			\centering
		\includegraphics[width=1\textwidth]{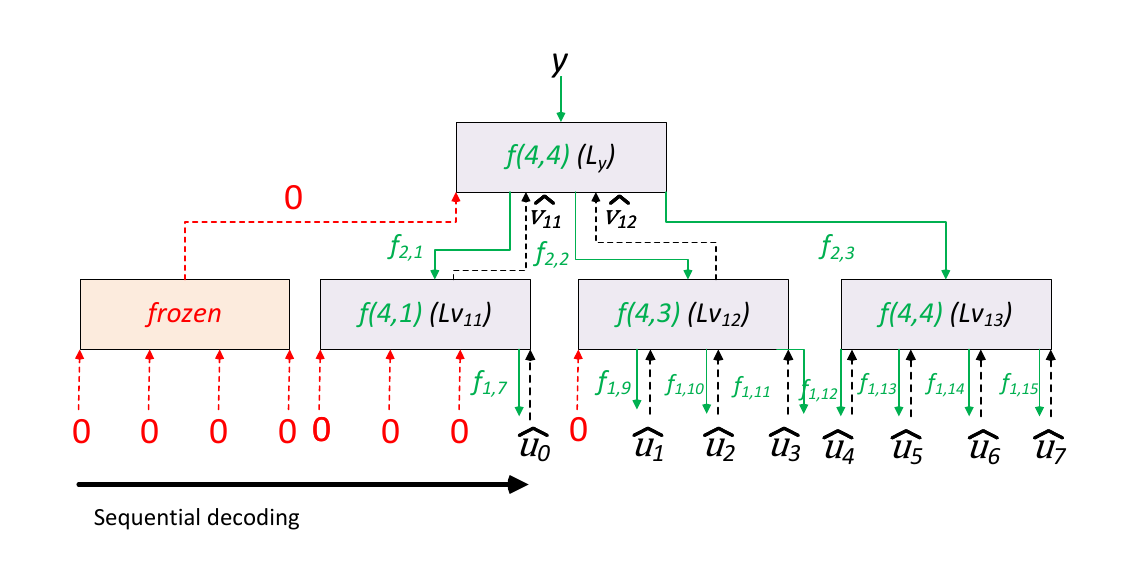}
		\caption{}
		\label{fig:decodertree}
	\end{subfigure}
	\centering
	\begin{subfigure}[t]{0.49\textwidth} 
		\centering
		\includegraphics[width=1\textwidth]{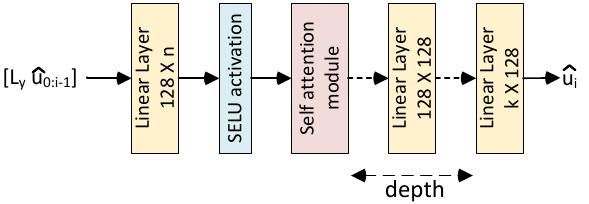}
		\caption{}
		\label{fig:decoderkernel}
	\end{subfigure}
	\hfill
	\centering
	\begin{subfigure}[t]{0.49\textwidth} 
		\centering
		\includegraphics[width=1\textwidth]{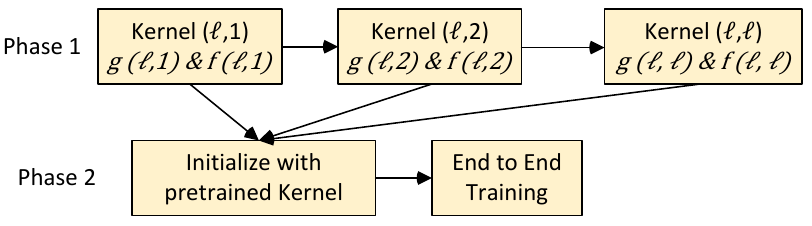}
		\caption{}
		\label{fig:curriculum}
	\end{subfigure}
	\caption{\small DeepPolar code architecture and training framework: (a) Hierarchical encoding structure for (16,8) code using 4×4 kernels, where each node $g_{d,b}$ represents a learned neural kernel at depth $d$ and bit position $b$ transforming input bits through a multi-layer network. (b) Sequential decoding structure with neural kernels $f_{d,b}$ maintaining the successive cancellation strategy while leveraging both channel information and previously decoded bits. (c) Multi-layer architecture of the DeepPolar+ decoder network $f_{d,\ell b+j}$ featuring self-attention mechanisms and feed-forward layers for enhanced bit-position dependency modeling. (d) Two-phase curriculum learning strategy where Phase 1 progressively trains individual kernels, followed by Phase 2 initializing the full code with pre-trained kernels for end-to-end optimization.}
	\label{fig: architecture and training}
\end{figure*}

\section{Background: DeepPolar Architecture \cite{hebbar2024deeppolar}}\label{Sec: background}
DeepPolar codes generalize polar codes by combining their structured encoding and decoding framework with learnable neural networks. A DeepPolar$(n,k,\ell)$ code consists of an encoder $g_{\phi}: \{0,1\}^k \rightarrow \mathbb{R}^n$ and a decoder $f_{\theta}: \mathbb{R}^n \rightarrow \{0,1\}^k$, where $\ell$ represents the neural kernel size. The encoding process starts with a message vector $\mathbf{m} \triangleq (m_1,\ldots,m_n) = (0,\ldots,u_0,0,\ldots,u_1,0,\ldots,u_{k-1},0,\ldots) \in \{0,1\}^n$ containing information bits $\mathbf{u}$ at indices corresponding to information set $\mathcal{I}$, with complement $\mathcal{F} = \mathcal{I}^C$ representing frozen positions.
\subsection{Encoding Architecture}
As shown in Fig.~\ref{fig:encodertree}, each encoder kernel $g_{d,b}$ at depth $d$ and position $b$ combines a deep neural network with polar-inspired elements:
\begin{equation}
	\begin{split}
		g_{d,b}(\mathbf{y}) = & \mathbf{W}^{(d,b)}_{L+1}\Big(\phi(\mathbf{W}^{(d,b)}_L\phi(\mathbf{W}^{(d,b)}_{L-1}\cdots\\
		&\mbox{\hspace{0.5cm}}\cdots\phi(\mathbf{W}^{(d,b)}_1\mathbf{y}))) + \mathbf{s}(\mathbf{y})\Big) + \alpha\mathbf{a}(\mathbf{y})
	\end{split}
\end{equation}
where $\mathbf{W}^{(d,b)}_i \in \mathbb{R}^{h \times h}$ are learnable weight matrices, with input layer $\mathbf{W}^{(d,b)}_1 \in \mathbb{R}^{h \times \ell}$ and output layer $\mathbf{W}^{(d,b)}_{L+1} \in \mathbb{R}^{\ell \times h}$. $\phi(\cdot)$ denotes the SELU activation function, and $\alpha$ is a binary flag for polar augmentation.
The polar augmentation $\mathbf{a}(\mathbf{y})= G^{\otimes \ell}\mathbf{y}$ preserves polarization through $\ell$-fold Kronecker products of the base polar kernel $G$. The skip connection pathway $\mathbf{s}(\mathbf{y})$ enhances gradient flow:
\begin{equation}
	\mathbf{s}(\mathbf{y}) = \mathbf{V}_M\phi(\mathbf{V}_{M-1}\phi(\cdots\phi(\mathbf{V}_1[\mathbf{y}; \mathbf{a}(\mathbf{y})])))
\end{equation}
where $\mathbf{V}_1 \in \mathbb{R}^{h \times 2\ell}$ is the input layer, $\mathbf{V}_i \in \mathbb{R}^{h \times h}$ are hidden layer weights for $i=2,\ldots,M$, with network depth $L=3$ and hidden dimension $h=64$.

\subsection{Decoding Architecture}
The decoder implements a neural successive cancellation strategy (Fig.~\ref{fig:decodertree}). Each component network $f_{d,\ell b+j}$ processes bits sequentially:

\begin{equation}
	\begin{split}
		f_{d,\ell b+j}(\mathbf{L}_y, \hat{\mathbf{u}}_{0:j-1}) = &\mathbf{U}^{(d,j)}_{L+1} \phi(\mathbf{U}^{(d,j)}_{L}\phi(\mathbf{U}^{(d,j)}_{L-1}\cdots\\
		&\mbox{\hspace{0.5cm}}\cdots \phi(\mathbf{U}^{(d,j)}_1[\mathbf{L}_y; \hat{\mathbf{u}}_{0:j-1}])))
	\end{split}
\end{equation}

where $\mathbf{L}y \in \mathbb{R}^\ell$ are channel log-likelihood ratios and $\hat{\mathbf{u}}_{0:j-1}$ represents previously decoded bits. The network comprises input layer $\mathbf{U}^{(d,j)}_1 \in \mathbb{R}^{h \times (\ell+j)}$, hidden layers $\mathbf{U}^{(d,j)}_i \in \mathbb{R}^{h \times h}$, and output layer $\mathbf{U}^{(d,j)}_{L+1} \in \mathbb{R}^{1 \times h}$, with depth $L=3$ and hidden dimension $h=128$.

The decoding follows:

\begin{equation}
	\hat{u}_i = \begin{cases}
		0 & \text{if } i \in \mathcal{F}\\
		0 & \text{if } i \in \mathcal{I} \text{ and } f_{d,\ell b+i}(\mathbf{L}_y, \hat{\mathbf{u}}_{0:i-1}) \geq 0\\
		1 & \text{if } i \in \mathcal{I} \text{ and } f_{d,\ell b+i}(\mathbf{L}_y, \hat{\mathbf{u}}_{0:i-1}) < 0
	\end{cases}
\end{equation}

\subsection{Training Methodology}
As illustrated in Fig.~\ref{fig:curriculum}, training follows a two-phase curriculum:
\begin{enumerate}
	\item Progressive kernel training from $(n=\ell, k=1)$ to $(n=\ell, k=\ell)$.
	\item Full code training using pretrained kernels and binary cross-entropy loss
\end{enumerate}

This approach enables efficient learning of effective coding strategies while maintaining the structural benefits of polar codes.

\section{DeepPolar+: Enhanced Architecture}\label{Sec: deepolar+}
Building on the DeepPolar framework, DeepPolar+ enhances both encoder and decoder architectures through self-attention mechanisms \cite{vaswani2017attention} and structured loss functions. While the encoder $g_{\phi}^+: \{0,1\}^k \rightarrow \mathbb{R}^n$ remains unchanged, the DeepPolar+ decoder $f_{\theta}^+: \mathbb{R}^n \rightarrow \{0,1\}^k$ extends the base decoder through attention-enhanced sequential processing.

\subsection{Attention-Enhanced Decoder}
For each kernel of size $\ell$, the component network $f^+_{d,\ell b+j}$ at depth $d$ and relative position $j$ ($j=0,1,\ldots,\ell-1$) from base position $b$ is parameterized as:

\begin{equation}
	\begin{split}
		f^+_{d,\ell b+j}(\mathbf{L}_y, \hat{\mathbf{u}}_{0:j-1}) = &\mathbf{\Theta}^{(d,j)}_{L+1} \phi(\mathbf{\Theta}^{(d,j)}_{L}\phi(\mathbf{\Theta}^{(d,j)}_{L-1}\cdots \phi(\\
		&\text{SA}(\phi(\mathbf{\Theta}^{(d,j)}_1[\mathbf{L}_y; \hat{\mathbf{u}}_{0:j-1}])))))
	\end{split}
\end{equation}

where the self-attention based mechanism SA($\cdot$) is defined as:

\begin{equation}
	\text{SA}(\mathbf{X}) = \text{LayerNorm}(\mathbf{X} + \text{Dropout}(\text{MultiHead}(\mathbf{X}, \mathbf{X}, \mathbf{X})))
\end{equation}

The multi-head attention operation processes input features through:

\begin{equation}
	\text{MultiHead}(\mathbf{Q}, \mathbf{K}, \mathbf{V}) = [\text{head}_1; \cdots; \text{head}_{h_a}]\mathbf{\Psi}^O
\end{equation}

where each attention head computes:

\begin{equation}
	\text{head}_i = \text{Attention}(\mathbf{Q}\mathbf{\Psi}^Q_i, \mathbf{K}\mathbf{\Psi}^K_i, \mathbf{V}\mathbf{\Psi}^V_i)
\end{equation}

\begin{equation}
	\text{Attention}(\mathbf{Q}, \mathbf{K}, \mathbf{V}) = \text{softmax}\left(\frac{\mathbf{Q}\mathbf{K}^T}{\sqrt{d_k}}\right)\mathbf{V}
\end{equation}
where $\mathbf{\Psi}^Q_i, \mathbf{\Psi}^K_i, \mathbf{\Psi}^V_i \in \mathbb{R}^{h \times d_k}$ are the query/key/value projections and $\mathbf{\Psi}^O \in \mathbb{R}^{h_ad_k \times h}$ is the output projection. The architecture uses $h_a=4$ attention heads with per-head dimension $d_k=32$ as empirically determined to balance computational efficiency with modeling capacity. Layer normalization and dropout (rate=0.1) provide regularization. The decoder uses network depth of $L=3$ with hidden dimension of $h=128$. 
The enhanced architecture offers several key advantages:
\begin{itemize}
	\item Dynamic feature interaction through self-attention enables sophisticated modeling of bit-position dependencies
	\item Layer normalization provides improved training stability
	\item Residual connections facilitate effective gradient flow
	\item Balanced attention head configuration optimizes computational efficiency and representational power
\end{itemize}

The decoder maintains the sequential decision rule of DeepPolar while leveraging these enhancements to learn more effective decoding strategies. The architecture alternates between:
\begin{enumerate}
	\item Multi-head attention layers for capturing global dependencies
	\item Position-wise feed-forward networks for processing local features
\end{enumerate}


\subsection{Enhanced Training Objective}
DeepPolar+ introduces a specialized loss function combining standard binary cross-entropy (BCE) with a novel soft block error rate term:
\begin{equation}
	\mathcal{L}_{\text{total}} = \mathcal{L}_{\text{BCE}} + \mathcal{L}_{\text{block}}
\end{equation}
where $\mathcal{L}_{\text{block}}$ is a differentiable block error objective:
\begin{equation}
	\mathcal{L}_{\text{block}} = -\frac{1}{B}\sum_{i=1}^B \log\left(\prod_{j=1}^k P(u_{i,j}|\hat{L}_{i,j}) + \epsilon\right)
\end{equation}
where $\hat{L}_{i,j}$ is the predicted log-likelihood ratio for bit $j$ in block $i$, $B$ is the batch size,  $P(u_{i,j}|\hat{L}_{i,j}) = \sigma(\hat{L}_{i,j})$ when $u_{i,j}=1$ and $1-\sigma(\hat{L}_{i,j})$ when $u_{i,j}=0$, $\sigma(x)$ is the sigmoid function,  and a small number $\epsilon=10^{-10}$ is chosen to prevent numerical underflow while minimally impacting the loss calculation. The model employs cosine annealing with warm restarts \cite{loshchilov2017sgdr} for learning rate scheduling.
This enhanced architecture balances both bit-level accuracy and block-level reliability while enabling effective training through improved gradient flow and optimization dynamics.
\begin{figure*}[t]
	\centering
	\begin{subfigure}[t]{0.49\textwidth} 
		\centering
		\includegraphics[width=1\textwidth]{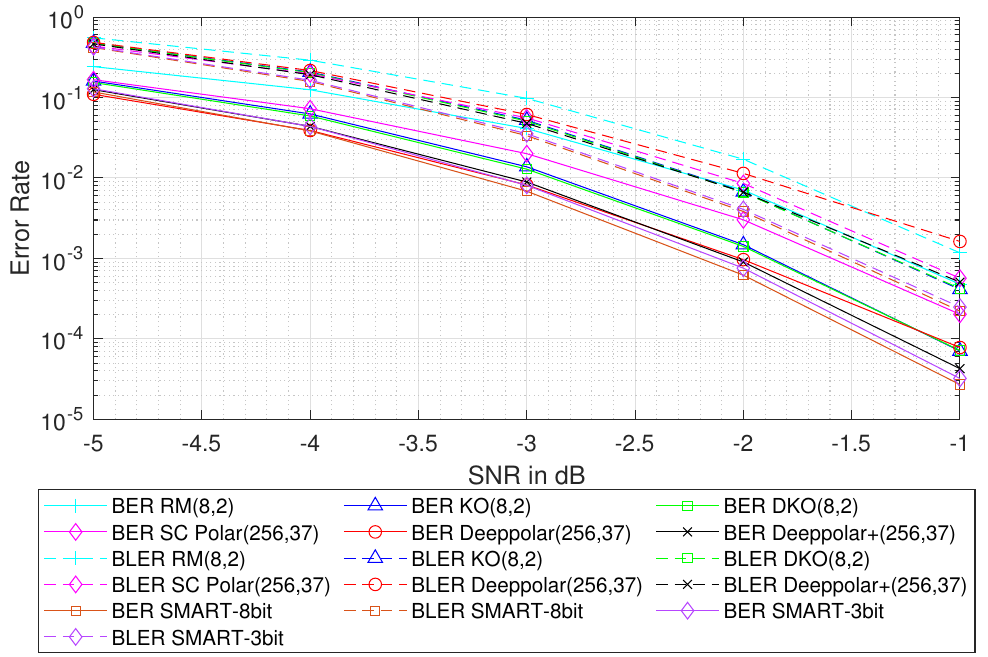}
		\caption{}
		\label{fig:smart_comparison}
	\end{subfigure}
	\hfill
	\centering
	\begin{subfigure}[t]{0.48\textwidth} 
		\centering
		\includegraphics[width=1\textwidth]{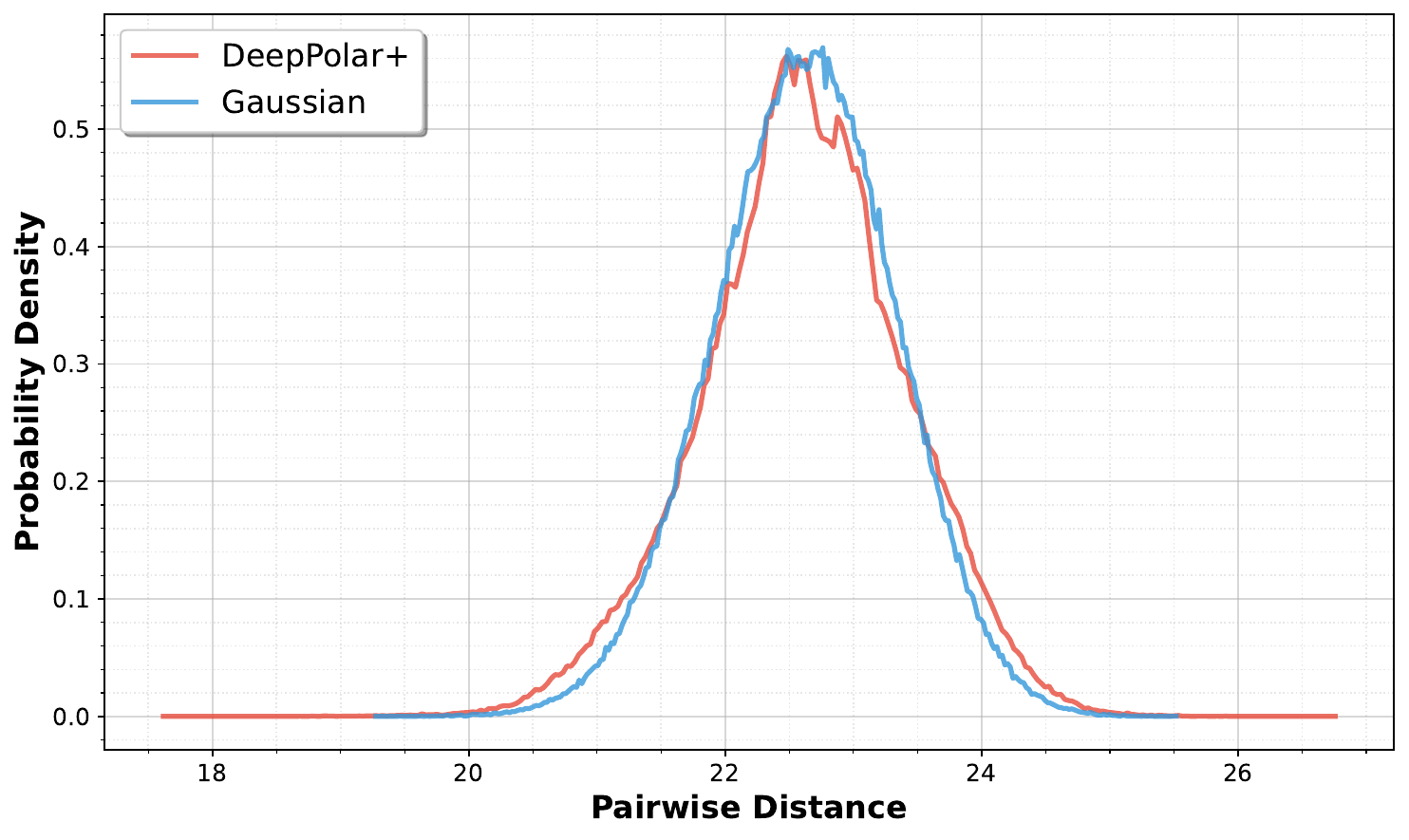}
		\caption{}
		
		\label{fig:pairwise}
	\end{subfigure}
	
	\centering
	\begin{subfigure}[t]{0.49\textwidth} 
		\centering
		\includegraphics[width=1\textwidth]{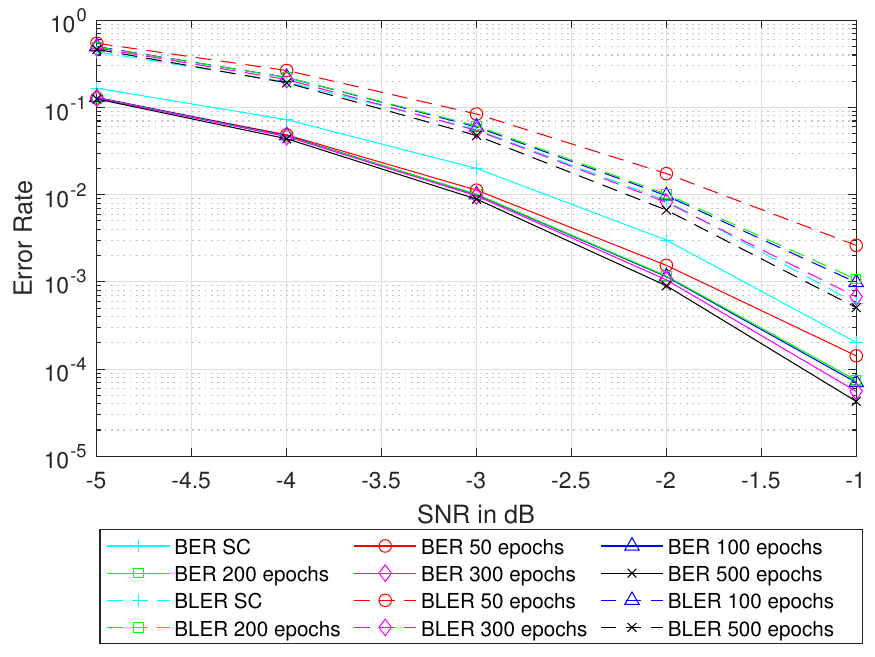}
		\caption{}
		\label{fig:convergence}
	\end{subfigure}
	\hfill
	\centering
	\begin{subfigure}[t]{0.49\textwidth} 
		\centering
		\includegraphics[width=1\textwidth]{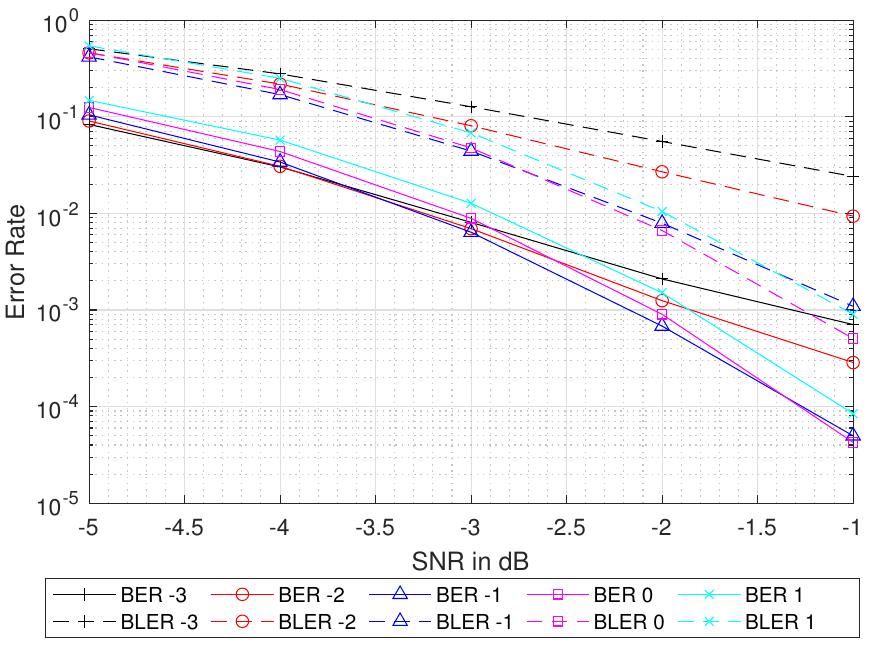}
		\caption{}
		\label{fig:difftrainingsnr}
	\end{subfigure}
	
	\caption{\small Comprehensive performance analysis of DeepPolar+ and its variants: (a) Performance comparison on AWGN channel showing BER (solid) and BLER (dashed) for DeepPolar+ and DeepPolar+SMART with 3-bit and 8-bit CRC against baseline approaches for (256,37) code, demonstrating consistent improvements over conventional SC decoding and DeepPolar.  (b) Pairwise Euclidean distance distribution analysis comparing DeepPolar+ codewords with random Gaussian codebook, revealing near-optimal spatial distribution. (c) Convergence analysis showing BER and BLER performance across different training durations, demonstrating efficient learning dynamics. (d) Impact of training SNR pairs on model performance, with (0,-2) dB configuration achieving optimal balance across the SNR range.}
	\label{fig: allberplots}
	
\end{figure*}
\subsection{DP+SMART Decoder: SNR-Matched Redundancy Technique for decoding DeepPolar+}
Analysis from Fig.~\ref{fig:difftrainingsnr} reveals that DeepPolar+ models exhibit SNR-dependent behavior, performing optimally near their training SNR but degrading at other SNR levels. Lower SNR-trained models excel in low SNR regimes but show error floors at higher SNRs, while the opposite holds for higher SNR-trained models. DP+SMART decoder addresses this through a novel framework combining multiple SNR-specialized models:\\
\textbf{CRC-Enhanced Encoding}: The system augments each input message $\mathbf{u} \in {0,1}^k$ with $r$ cyclic redundancy check (CRC) bits before encoding:
\begin{equation}
	\mathbf{x} = g_{\phi}^+([\mathbf{u}, \text{CRC}(\mathbf{u})]) \in \mathbb{R}^n
\end{equation}
\textbf{Multi-Model Parallel Decoding}: Upon receiving the corrupted signal $\mathbf{y}$ through the communication channel, the system employs an ensemble of $M$ specialized decoders:
\begin{equation}
	\mathcal{D} = \{\hat{\mathbf{u}}_i = f_{\theta_i}^+(\mathbf{y})\}_{i=1}^M
\end{equation}
where each decoder $f_{\theta_i}^+$ is optimized for a specific SNR regime.\\
\textbf{CRC-Guided Selection}: The system determines the final decoded message through an adaptive selection mechanism:
\begin{equation}
	\hat{\mathbf{u}} = \begin{cases}
		\hat{\mathbf{u}}_k & \text{if } \exists k: \text{CRC}(\hat{\mathbf{u}}_k) \text{ valid}\\
		f_{\theta_b}^+(\mathbf{y}) & \text{otherwise}
	\end{cases}
\end{equation}
where $f_{\theta_b}^+$ represents a baseline decoder trained for robust general performance.
This framework enables efficient parallel implementation with CRC-based reliability validation, achieving superior performance across diverse channel conditions.
%
%
%

\section{Results}\label{Sec:results}
We conduct extensive experiments to evaluate DeepPolar+ focusing on both BER and BLER performance across various SNRs (1/$\sigma^2$) in AWGN channel with noise variance $\sigma^2$ .  We have compared the performance with conventional polar codes with SC decoding, DeepPolar, Reed-Muller codes with Dumer decoding\cite{1057465} \cite{6499441}, and KO \cite{pmlr-v139-makkuva21a} /DKO \cite{10619329} codes.

\subsection{Experimental Setup}
We evaluate codes with block length $N=256$ and information length $K=37$, implemented in PyTorch with the following configuration:
\begin{itemize}
	\item Kernel size $\ell = 16$ (empirically determined as $\sqrt{N}$)
	\item Training SNR range: -5 dB to 0 dB 
	\item Batch size: 20,000 examples
	\item Optimizer: Adam \cite{DBLP:journals/corr/KingmaB14} with learning rate $10^{-3}$
	\item Training: 500 epochs with cosine learning rate scheduling
\end{itemize}

\subsection{Performance Analysis}
Our experimental results in Fig. \ref{fig:smart_comparison} demonstrate that DeepPolar+ achieves substantial improvements in both BER and BLER performance. At BER=$10^{-4}$, DeepPolar+ provides a 0.4 dB gain over conventional SC decoding and a 0.2 dB improvement over the original DeepPolar. Crucially, DeepPolar+ resolves the BER-BLER trade-off limitation, matching SC decoding's block error performance (slightly exceeds the SC decoding's BLER performance by 0.08 dB for $10^{-3}$ BLER) while maintaining superior bit error performance.
The enhanced architecture demonstrates improved training efficiency, requiring only 500 epochs compared to DeepPolar's 2000 epochs - a 75\% reduction attributed to:
\begin{itemize}
	\item Self-attention mechanisms capturing long-range dependencies
	\item Structured loss function jointly optimizing BER and BLER
	\item Effective exploration of the loss landscape through cosine learning rate scheduling
\end{itemize}

\subsection{Analysis of Codeword Distance Distribution}
To understand the learned coding strategy, we analyze the pairwise Euclidean distances between codewords, sampling 10,000 codewords and normalizing by code length N=256. As shown in Fig. \ref{fig:pairwise}, DeepPolar+ generates codewords with a distance distribution closely resembling that of random Gaussian codebooks - known to be capacity-achieving asymptotically. This suggests DeepPolar+ effectively learns to utilize the available signal space while maintaining beneficial polar code properties.

\subsection{Convergence Analysis} 
Fig. \ref{fig:convergence} shows BER and BLER performance across different training durations (50-500 epochs). The results reveal:
\begin{itemize}
	\item Most significant BER improvements occur within 200 epochs
	\item BLER performance requires extended training, particularly at higher SNRs
	\item Convergence rate varies with SNR region - faster at lower SNRs (-5 to -3 dB), slower at higher SNRs where error events are sparser
\end{itemize}

\subsection{DP+SMART Performance}
DP+SMART combines models trained on five SNR pairs: (0,-2), (-1,-3), (-3,-5), (1,-1), and (-2,-4) dB, with CRC verification using standard 3-bit and 8-bit polynomials. The (0,-2) dB configuration serves as both the default encoder model and the fallback decoder when no CRC matches are found. The system incorporates CRC verification using two standard polynomials:
\begin{equation}
	g_3(x) = 1 + x + x^3  \text{ (3-bit CRC)}
\end{equation}
\begin{equation}
	g_8(x) = 1 + x + x^3 + x^5 + x^8  \text{ (8-bit CRC)}
\end{equation}. 

As shown in Fig. \ref{fig:smart_comparison}, at a BER of $10^{-4}$, DP+SMART decoder with 8-bit CRC achieves gains of approximately 0.15 dB over DeepPolar+, 0.33 dB over DeepPolar, and 0.5 dB over SC decoding. The improvements in BLER performance are even more pronounced, showing around 0.33 dB gain over SC decoding and 0.6 dB gain over DeepPolar at a BLER of $10^{-3}$.
The 3-bit CRC variant offers an attractive compromise between performance and overhead. The performance gaps between 3-bit and 8-bit CRC variants are relatively small: 0.07 dB at BER of $10^{-4}$ and 0.04 dB at BLER of $10^{-3}$. Both CRC variants show increasing performance gains at higher SNRs (-2 to -1 dB), demonstrating the effectiveness of CRC-guided model selection in leveraging SNR-specific decoders. Notably, DeepPolar+SMART maintains consistent performance across the entire SNR range, avoiding the performance degradation typically observed in single-model approaches operating outside their training SNR.
These results establish DeepPolar+ with DP+SMART decoder as a robust solution for practical implementation of neural polar codes, effectively addressing both performance and reliability requirements across diverse channel conditions.

 While the current work adopts a successive cancellation decoder inspired architecture for decoder, DeepPolar+ can be naturally extended to list decoding by maintaining L parallel decoding paths. This DeepPolar+List extension would explore multiple candidate paths simultaneously, potentially offering further improvements in error-correction performance at high SNRs - a promising direction for future research.

\section{Conclusion}\label{Sec: conclusion}
This paper introduces DeepPolar+, a significant advancement in neural polar coding that effectively resolves the BER-BLER trade-off limitation of previous approaches. Through attention-enhanced decoding, structured loss functions, and adaptive SNR-matched transmission, DeepPolar+ achieves substantial performance improvements for the challenging (256,37) code configuration - demonstrating a 0.4 dB gain in BER at $10^{-4}$ compared to conventional SC decoding while matching its BLER performance.
The DeepPolar+ with DP+SMART decoder framework further extends these gains, achieving up to 0.5 dB improvement in BER at $10^{-4}$ and 0.6 dB in BLER at $10^{-3}$ over the original DeepPolar architecture. Beyond raw performance metrics, our approach offers practical benefits through a 75\% reduction in training epochs. The close correspondence between DeepPolar+'s learned codeword distribution and theoretical Gaussian codebooks suggests near-optimal coding strategies while maintaining polar code structure.
 DeepPolar+ represents a small step toward practical neural polar codes, effectively bridging the gap between theoretical potential and implementation requirements.



\end{document}